\documentclass[iop,apj]{emulateapj}
\usepackage{float}

\newcommand{\msun}{$M_\odot$}
\newcommand{\kms}{km s$^{-1}$~}
\newcommand{\ugp}{$u'g'r'i'z'$~}

\def\arcmin{$^{\prime}$~}
\def\arcsec{$^{\prime\prime}$~}
\def\gapp{{_>\atop{^\sim}}}
\def\lapp{{_<\atop{^\sim}}}

\shorttitle{Searching for Optical Counterparts to HI Clouds}
\shortauthors{Janesh et al.}

\begin{document}

\title{Searching for Optical Counterparts to Ultra-compact High
  Velocity Clouds: Possible Detection of a Counterpart to
  AGC~198606} 
  
\author{William Janesh$^1$, Katherine
  L. Rhode$^1$, John J. Salzer$^1$, Steven
  Janowiecki$^1$, Elizabeth A. K. Adams$^2$,
  Martha P. Haynes$^3$, Riccardo Giovanelli$^3$,
  John M. Cannon$^4$, Ricardo R. Mu\~noz$^5$}

\affil{$^1$
Department of Astronomy, Indiana University, 727 E. Third Street,
Bloomington, IN 47405, USA
}
\affil{$^2$
Netherlands Institute for Radio Astronomy (ASTRON), Postbus 2, 7900 AA
Dwingeloo, The Netherlands 
}
\affil{$^3$
Center for Radiophysics and Space Research, Space Sciences Building, Cornell University, Ithaca, NY 14853, USA 
}
\affil{$^4$
Department of Physics and Astronomy, Macalaster College, 1600 Grand
Avenue, Saint Paul, MN 55105, USA 
}
\affil{$^5$
Departamento de Astronomía, Universidad de Chile, Casilla 36-D,
Santiago, Chile
}
\email{wjanesh@indiana.edu}
\begin{abstract}
We report on initial results from a campaign to obtain optical imaging
of a sample of Ultra Compact High Velocity Clouds (UCHVCs) discovered
by the ALFALFA neutral hydrogen (HI) survey.  UCHVCs are ALFALFA
sources with velocities and sizes consistent with their being low-mass
dwarf galaxies in the Local Volume, but without identified optical
counterparts in existing catalogs.  We are using the WIYN 3.5-m
telescope and pODI camera to image these objects and search for the
presence of an associated stellar population.  In this paper, we
present our observational strategy and method for searching for
resolved stellar counterparts to the UCHVCs.  We combine careful
photometric measurements, a color-magnitude filter, and spatial
smoothing techniques to search for stellar overdensities in the 
$g$- and $i$-band pODI images.  We also run statistical tests to
quantify the likelihood that whatever overdensities we find are real
and not chance superpositions of sources.  We demonstrate the method
by applying it to two specific data sets: WIYN imaging of Leo~P, a
UCHVC discovered by ALFALFA and subsequently shown to be a low-mass
star-forming dwarf galaxy in the Local Volume; and WIYN imaging of
AGC198606, an HI cloud identified by ALFALFA that is near in position
and velocity to the Local Group dwarf galaxy Leo~T.  Applying the
search method to the Leo~P imaging data yields an unambiguous
detection ($>$99\% confidence) of this galaxy's stellar population.
Applying our method to the AGC198606 imaging yields a possible
detection (92\% confidence) of an optical counterpart located $\sim2.5$ arc
minutes away from the centroid of AGC198606's HI distribution but still within the HI disk. We use
the optical data to estimate a distance to the stellar counterpart between
373 and 393~kpc, with an absolute magnitude of $M_i = -4.67 \pm 0.09$.
Combining the WIYN data with our previous estimate of the HI mass of
AGC198606 from WSRT imaging yields an HI-to-stellar mass ratio of
$\sim$45$-$110.
\end{abstract}


\section{Introduction}
\label{section:introduction}
A current issue in cosmology is the disagreement between models and
observations regarding the number and mass distribution of low-mass
galaxies that exist in environments like the Local Group.  Simulations
of the formation of structures like the Local Group in a $\Lambda$CDM
universe predict large numbers of low-mass dark matter halos around
the Milky Way and M31 \citep[e.g.,][]{moore06,diemand07,gk14}.
Observational campaigns aimed at searching for low-mass galaxies in our local neighborhood,
while successful at detecting some new Local Group dwarf galaxies 
\citep[e.g.,][]{willman05,martin13,2015arXiv150302584T} have nevertheless not yet
found them in sufficient numbers to match the models; this is the
so-called ``missing satellites'' problem \citep{kauffmann93,klypin99,moore99}.

Some of the discrepancies between observational and theoretical results can be mitigated when internal
evolutionary processes (e.g., ``feedback'' from star formation and
supernovae) and external processes (e.g., reionization, tidal
stripping, ram pressure stripping) are taken into account. The details of how and when these processes occur, and what effect they have on the galaxies, are complicated \citep[e.g.,][]{bullock00,guo10,papastergis12}. The net effect, however, may be that the cold baryons in nascent galaxies become so depleted that it is impossible for the galaxies to form stars.
Alternatively, feedback may limit the star formation that does occur to only one
generation of stars, rendering any stellar population that is present
undetectable given our current observational capabilities and the
limitations of the surveys completed so far.

In order to both explore the ``missing satellites'' problem and to
investigate the complex interplay between low-mass dark matter halos,
their baryonic components, and the gas and stellar processes that
shape them, we have initiated a project to obtain optical observations
of a specific class of candidate low-mass galaxies discovered via
their neutral hydrogen emission.  These objects were identified by the
Arecibo Legacy Fast ALFA (ALFALFA) neutral hydrogen (HI) survey
\citep{alfalfa}.  The ALFALFA survey uses data from the Arecibo
305-m telescope to map 7000 deg$^2$ of the sky in the HI 21 cm
line at an angular resolution of $\sim 3'.5$ and a spectral
resolution of $\sim$5 km~s$^{-1}$.
Data acquisition for the ALFALFA survey was completed in October 2012;
full processing of all the survey data cubes has been concluded, and
catalogs have been produced for all sources at $cz$ $<$
3000~km~s$^{-1}$.  ALFALFA has detected a sample of isolated
ultra-compact high-velocity HI clouds \citep[UCHVCs;][]{giovanelli10,adams13} with kinematic
properties that make them likely to be located within the Local
Volume, but not part of the Milky Way. The selection of these UCHVCs
is described in \citet{adams13}; briefly, sources included in the
sample must have $|v_{LSR}| > 120$ \kms to exclude Galactic HI
sources, a diameter less than 30\arcmin in size, and a signal-to-noise
ratio $\geq 8$. Sources must also be well-isolated from larger HI
structures and previously known high-velocity clouds. Notably, while they are compact, 
the UCHVCs are generally spatially resolved. The members of the UCHVC
sample possess HI properties similar to those of Leo~T
\citep{irwin07}, which is the faintest known Local Group dwarf galaxy with
evidence of recent star formation ($M_V \simeq -7$). At a distance of 1
Mpc, the UCHVCs would have HI masses of $\sim 10^5-10^6$ \msun, HI
diameters of $\sim2-3$ kpc, and dynamical masses of $\sim 10^7-10^8$
\msun, placing them in the range of lower mass dark matter halos. We
are using the WIYN Observatory\footnote{The WIYN Observatory is
  a joint facility of the University of Wisconsin-Madison, Indiana
  University, the University of Missouri, and the National Optical
  Astronomy Observatory.}  to image the UCHVCs in order to investigate
whether these objects are indeed nearby dwarf galaxies and, if
possible, to study their optical emission and stellar populations.

A dwarf galaxy located just outside the Local Group was recently identified using this type of approach: Leo P, a gas-rich, star-forming dwarf galaxy with $M_V \sim -9.4$ and a distance of $\sim 1.7$ Mpc, was discovered via its ALFALFA HI detection and follow-up optical imaging with WIYN \citep{leop1,rhode13}. Several other studies have employed similar strategies to look for nearby dwarfs.  For example, \citet{tollerud15} targeted HI sources found in the GALFA HI survey and identified two objects in WIYN follow-up imaging that may be Local Volume dwarf galaxies. \citet{secco} used deep imaging with the Large Binocular Telescope to search for optical counterparts to a subset of the \citet{adams13} UCHVCs and found a distant low surface brightness counterpart to an ALFALFA source \citep[see also][]{sand15,adams15b}.  This process -- i.e., HI source identification followed by optical imaging to identify a detectable stellar population and provide a distance estimate -- has thus been shown to be a potentially important path for finding more Local Volume dwarfs, for placing limits on how many nearby low-mass galaxies may remain undetected, and for understanding how such previously-unknown dwarf galaxies fit in with the rest of the nearby galaxy population \citep[e.g.][]{leop4}.

For the current project -- follow-up imaging of UCHVCs with WIYN -- we have devoted much of our 
initial effort to analyzing images of the ALFALFA source AGC198606 
\citep[see][and Table~\ref{table:properties}]{adams15a}.  Despite a
radial velocity below the threshold for inclusion in
the \citet{adams13} sample (51 km~s$^{-1}$), the object has similar properties to Leo~T's HI
component and is in fact near Leo~T in both position and velocity
(separations of $1^{\circ}.2$ and 16 \kms, respectively), earning it
the nickname ``Friend of Leo~T''.
\citet{adams15a} used the Westerbork Synthesis Radio Telescope (WSRT)
to obtain follow-up observations of AGC198606 and derived an HI mass
of $3.5 \times 10^6 d_{Mpc}^2$ \msun, an intrinsic velocity dispersion
of 9.3 \kms, a rotation velocity of $\sim14$ \kms, and a dynamical
mass of $3.5 \times 10^8 d_{Mpc}$ \msun.

In this paper we present our observational strategy, analysis
methods, and techniques for detecting stellar populations in the
optical follow-up images of the ALFALFA-detected UCHVCs.
We then demonstrate these detection methods by showing results from
WIYN imaging of two fields: AGC208583 (Leo~P) and AGC198606 (Friend of
Leo~T).
Not surprisingly, applying the method to the former field yields an
unambiguous detection
of the resolved stellar population of Leo~P.  Applying the method to
the AGC198606 field yields a possible detection of an optical
counterpart to this HI source, even though \citep[as noted in][]{adams15a} no conspicuous stellar counterpart is present in the images.

The paper is organized as follows: in Section \ref{section:techstuff}, we
describe the WIYN observations, image processing, source
detection, and photometric measurements.  Section \ref{section:filter}
presents the filtering and smoothing steps that are performed for each
set of images to search for optical counterparts to the ALFALFA
sources. Section \ref{section:examples} shows the results of this process for
the Leo~P field and the AGC198606 field.  The statistical significance
and the implications of the possible detection of a counterpart to
AGC198606 are also discussed.  The last section summarizes our main
conclusions and describes the current status of our optical follow-up
imaging campaign.
\section{Data Acquisition and Initial Processing}\label{section:techstuff}

\subsection{Observations and Image Reductions}
Our targets for optical follow-up are UCHVCs selected from
\citet{adams13} as well as other ALFALFA sources that satisfy similar
selection criteria but are not part of the original \citet{adams13} catalog. We also add additional UCHVC candidates as the ALFALFA catalog expands beyond the initial 40\% complete data release.  We observed the sources with the partially-filled
One Degree Imager (pODI) on the WIYN 3.5-m telescope at Kitt Peak
National Observatory\footnote{Kitt Peak National Observatory, part of
  the National Optical Astronomy Observatory, is operated by the
  Association of Universities for Research in Astronomy (AURA) under a
  cooperative agreement with the National Science Foundation.}. 
The pODI camera has a $\sim 24' \times 24'$ field of view and a pixel
scale of 0.11\arcsec per pixel on WIYN. For each target, nine 300-second exposures were executed in a preset dither pattern in both the SDSS $g$ and $i$ filters, with some
targets also observed in SDSS $r$. The majority of data collection for this project took place in March and April 2013, with
additional observations acquired in March, May, and October 2014. Future data collection will use an upgraded WIYN ODI camera with a $40' \times 48'$ field of view. Observations and data analysis are ongoing; this paper presents the result of our efforts to establish the image processing methods and analysis techniques for the project.
 The seeing, as characterized by the full width half maximum of the point spread function (FWHMPSF), in the images we have acquired so far ranges between 0.7\arcsec\ and 1.3\arcsec. The particular dataset presented in later sections is representative of both the image quality and depth of the pODI observations we have obtained to date. 

Raw images were transferred to the ODI Pipeline, Portal, and Archive
\citep[ODI-PPA][]{gopu14, young14} at Indiana University, then processed
using the QuickReduce data reduction pipeline \citep{kotulla14} to
remove the instrumental signature. The QuickReduce pipeline includes
standard bias, dark, and flat corrections, as well as a pupil ghost
correction, non-linearity corrections, cosmic-ray removal, and fringe
removal where necessary.  To account for image artifacts that remain after the QuickReduce processing, we constructed and applied an
illumination correction by creating a dark-sky flat derived from the science
observations. To create the flat on a filter-by-filter basis, we first
masked all objects in the images, combined the masked images
into a single image with a median algorithm, and then applied a smoothing kernel to
the combined image to construct
the final dark-sky flat field. We then divided each individual
pODI image by the appropriate dark-sky flat field. This illumination
correction was necessary in order to remove artifacts from
mis-matched sky levels between dithered images. The final result is very flat 
images over the full field-of-view of the camera \citep{janowiecki15}. 
The processed object images were then combined by scaling each image to a 
common flux level using Sloan Digital Sky Survey (SDSS) DR9 catalog stars 
\citep{ahn12} in the field, when available, to account for varying sky
transparency. Additionally, we used the DR9 catalog stars to determine
photometric solutions, and projected all individual images of the same
field to a common pixel scale defined by the SDSS stars' World
Coordinate System (WCS) solution. Finally, the processed object images
were stacked into a single combined image in each filter for each
UCHVC field.

\subsection{Source Detection and Photometry}

A common set of steps is executed on each of the final combined images.
First, the images are cropped in order to remove the higher-noise
edges of the dither pattern, leaving the central $\sim$20\arcmin\ by
20\arcmin of the pointing.  The mean standard deviation of the
background level is measured in $\sim$10$-$20 empty regions of the
image.  Using IRAF tasks \texttt{daofind} and \texttt{phot}, we
identify all sources with peak counts at least $4.0$ times the
background noise level and measure their instrumental magnitudes.  We 
explored the use of lower \texttt{daofind} thresholds but
found that lower detection thresholds typically produced a large number
of spurious sources without finding any additional genuine objects.
We then mask out saturated stars (with diffraction spikes and/or bleed
trails) and bright background galaxies and remove spurious sources
with undefined magnitudes.  To create a single list of sources
appearing in both filters, we match sources in the $g$ and $i$
images and then measure the full width at half maximum (FWHM) of the
radial profile of each matched source.  Obviously extended objects,
with large FWHM values in a plot of FWHM vs. instrumental magnitude,
are removed. The FWHM above which we remove objects varies as a function of the instrumental magnitude, but a representative value is 1.5 times the mean stellar FWHM derived from bright stars in the image. An aperture correction is determined for each image and
applied to the measured instrumental magnitudes of the remaining
sources.  We convert these instrumental magnitudes to calibrated
magnitudes by applying the zero points and color terms calculated from
the SDSS stars in the images. Galactic reddening corrections are
determined from \citet{sf11}\footnote{Galactic
  extinction values were obtained from the NASA/IPAC Infrared Science
  Archive at \url{http://irsa.ipac.caltech.edu/applications/DUST/}} and then
applied to produce the final set of magnitudes and colors.  At the end
of this process, the pODI images with good seeing (FWHMPSF $\lapp$ 0.8'')
typically yield $\gapp 2000$ sources and a $5\sigma$ limit on the
brightness of a point source of $\sim$25 magnitude in both $g$ and
$i$ filters.

\section{Detection Technique}\label{section:filter}
\subsection{Color-magnitude filter}

Our objective is to search for the resolved stellar population of the
optical counterpart for each UCHVC we observe.  We follow a similar
strategy to that outlined in \citet{wwj} and \citet{betseythesis},
making adjustments as appropriate for our data set.  The first step is
to apply a filter in
color-magnitude space to the calibrated photometry of the sources in
each field.

\begin{figure*}[ht!]
\begin{center}
\includegraphics[scale=0.8]{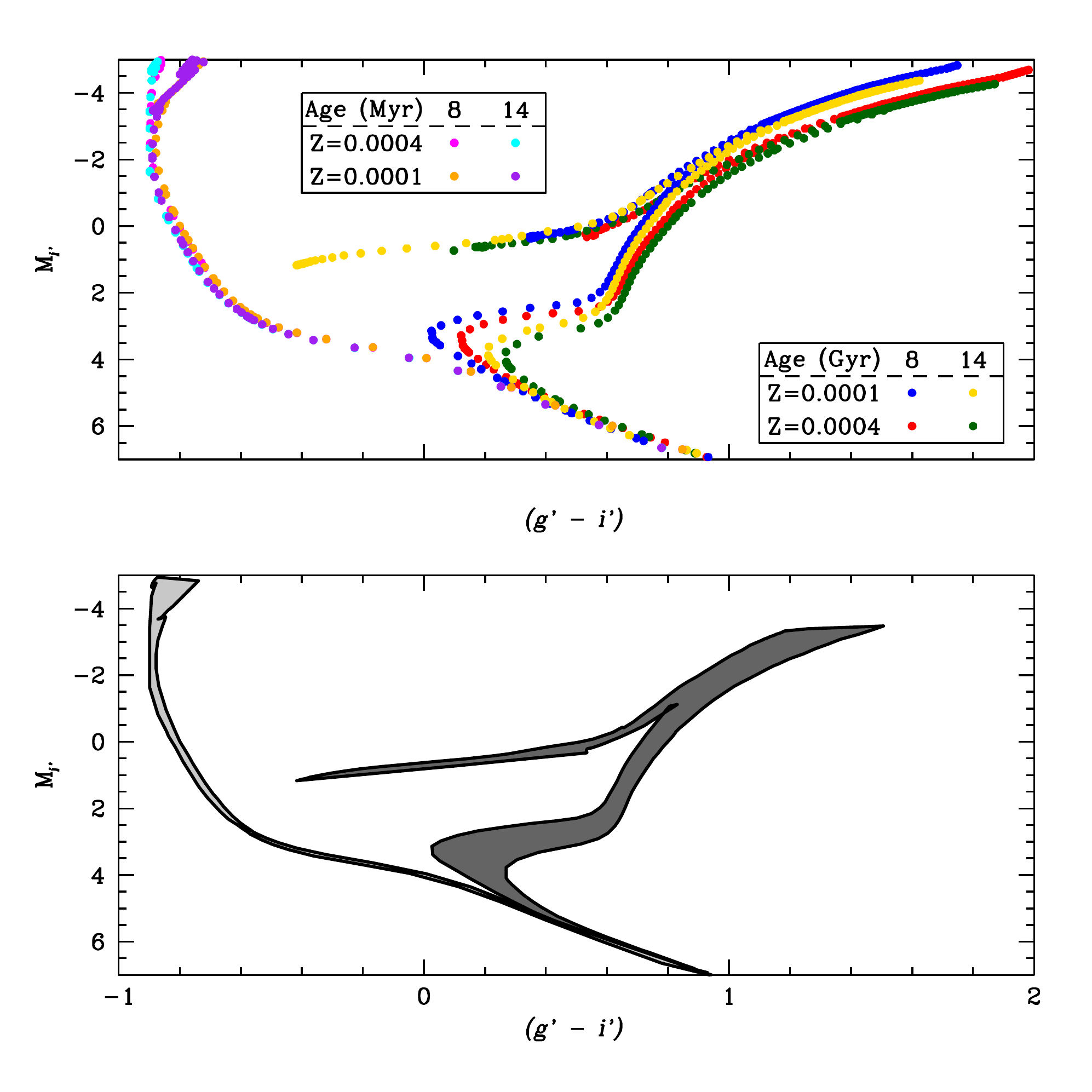}
\end{center}
\caption{Construction of the color-magnitude filter we use to select
  stars consistent with an old or young stellar population. The
  colored points show \citet{girardi} isochrones, and the shaded gray
  regions enclose our final filter boundaries.  Dark gray is the old
  stellar population, and light gray is the young stellar
  population. 
  Note that the old stellar population filter includes the region up to
  and including the tip of the Red Giant Branch, and does not include
  the Asymptotic Giant Branch region.\label{filterconstruct}}
\end{figure*}

To construct a color-magnitude diagram (CMD) filter, we select
\citet{girardi} isochrones for both old ($8 \leq t_{age}/$Gyr $\leq
14$) and young ($8 \leq t_{age}/$Myr $\leq 14$) stellar
populations. Because isolated HI clouds are unlikely to be highly chemically
enriched by the products of stellar evolution, we use the metallicity
range of $Z=0.0001$ to $Z=0.0004$ (1/200 to 1/50 of the solar value)
subsets for each age group to further define the filter.
The extent in color-magnitude space (i.e., in the $M_i$ vs. $g-i$
plane) covered by these isochrones is used to define the outer
boundaries of the CMD filter.  Figure \ref{filterconstruct} shows the
isochrones selected to construct the filter, as well as the final
CMD filter we adopt. We note that for some fields, such as the AGC198606 field (Section \ref{section:198606}), there is no evidence of bright, blue stars or ongoing star formation. In such cases we apply a CMD filter that includes only the old stellar population, because there are simply no stars in the region of the CMD where young stars would appear.

Before it is applied to the source photometry, the CMD filter is
projected to a given distance within the Local Volume by adding the
appropriate distance modulus to the $M_i$ value of the original
filter.  We shift the filter so as to sample the range of distances
over which we expect to be able to detect stellar counterparts to the
UCHVCs: from 250 kpc (the closest distance at which we expect to find 
dark matter halos with a significant HI counterpart; \citet{adams13}, \citet{spekkens14}) 
to 2.5 Mpc (the distance to the farthest observable
Red Giant Branch star in a typical image from our data set). In terms
of the distance modulus, we sample a range of $m-M$ values from 22.0
to 27.0 in steps of 0.4 magnitude. When we find a
potentially significant detection in the filtering/smoothing process,
we re-sample the range of $m-M$ values around which the detection
occurs with a more finely-spaced grid, in order to determine a more
precise distance for the putative counterpart.

The calibrated 
magnitude and 
color of each source is tested to determine whether the source falls
within the CMD filter that corresponds to a specific distance.
The source passes the test if its position in the $i$ versus $g-i$
plane falls within the boundaries of the CMD filter
or if the $1\sigma$ error bars (in either the magnitude or color
direction) for the source overlap the filter boundaries.
This has the effect of widening the CMD filter at fainter magnitudes.

\subsection{Spatial smoothing}\label{section:smooth}

Once stars have been selected with the CMD filter at a given distance,
we smooth the stars' spatial distribution on the expected scale of a
dwarf galaxy. This smoothing is achieved by first binning the
remaining filtered stars into a grid with bin size $\sim 7$'' $\times
\sim 7$''. We will refer to the bins in this grid as pixels, since in
essence they represent a low-resolution image of the field. We then
convolve each pixel in the grid with a Gaussian kernel with a size roughly
equivalent to the spatial scale we expect for a nearby dwarf galaxy
(FWHM $\simeq$ 2 arcmin).  This spatial scale is comparable to the
optical extent of the dwarf galaxies Leo~T \citep[half-light radius
  $r_h = 1.4$ arcmin;][]{irwin07} and Leo~P \citep[optical extent $\sim
  90$ arcsec;][]{rhode13}.  Pixels in the convolved image (which we
call $A$, following the convention of \citet{wwj}) with larger signal
correspond to areas of the WIYN image with higher surface densities of
stars that have passed the CMD filter criteria.  We test 
whether these pixels contain genuine overdensities by calculating the
mean signal level and its standard deviation over the entire $A$ image. We
then calculate a new image, $S$, such that

\begin{equation} \label{significance}
S(x,y) = \frac{A(x,y) - \bar{A}}{A_{\sigma}}
\end{equation}

\noindent where $\bar{A}$ and $A_{\sigma}$ are, respectively, the mean and
standard deviation of the pixel values in $A$.  This new image ($S$)
effectively gives us the strength of the signal in $A$ by measuring
the number of standard deviations a given pixel is above or below the
mean signal level, and provides an easy way to identify overdensities
in the field.

\begin{figure}[ht!]
\begin{center}
\includegraphics[scale=0.5]{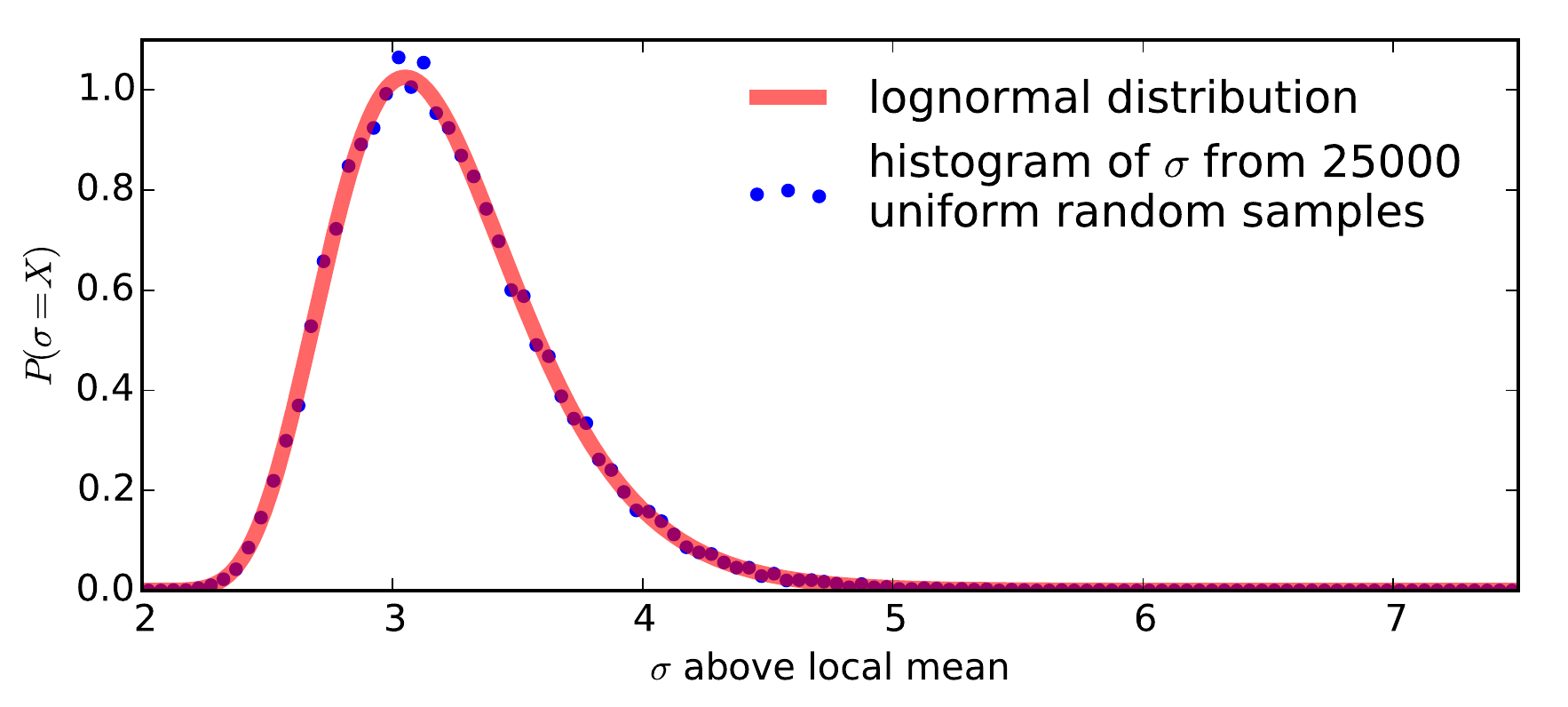}
\end{center}
\caption{Distribution of peak $\sigma$ values for 25000 random samples
  with 400 points uniformly distributed across a 20\arcmin by 20\arcmin field-of-view, which is well-fitted by a
  log-normal probability distribution function. \label{peakdist}}
\end{figure}

To evaluate the outcome of the filtering and smoothing process,
we 
implement a technique that allows us to quantify the likelihood
that an overdensity of a given strength (as defined in the $S$ image)
is a legitimate detection of a possible optical counterpart rather
than simply a stochastic variation. We use a Monte Carlo technique and
generate 25000 samples of uniformly-distributed random points in a
square 20\arcmin\ on a side, which is equal to the field-of-view of
our cropped WIYN pODI images. The number of objects in each of the
25000 realizations is set equal to the number of point sources that
pass the CMD filter criteria for a given field.
We then perform the smoothing steps described above for each random
realization and track the strength of the pixel with the highest
signal level in the resultant $S$ image. An example distribution of
these peak values for a sample size of 400 point sources is
shown in Figure \ref{peakdist}.

For each field and each application of the CMD filter, we can compare
the peak overdensity value derived from the real, observed image data
to the distribution of values generated from the random-sample data.
In this way we can estimate the statistical significance of a given
peak in the observed $S$ image and assign a robust confidence level to
detections.  This is especially useful for images in which a marked
overdensity is detected but which have no visually obvious optical
counterpart.

\section{Application of Technique}\label{section:examples}
\subsection{Leo P}\label{section:leop}

\begin{figure*}[ht!]
\begin{center}
\includegraphics[scale=0.8]{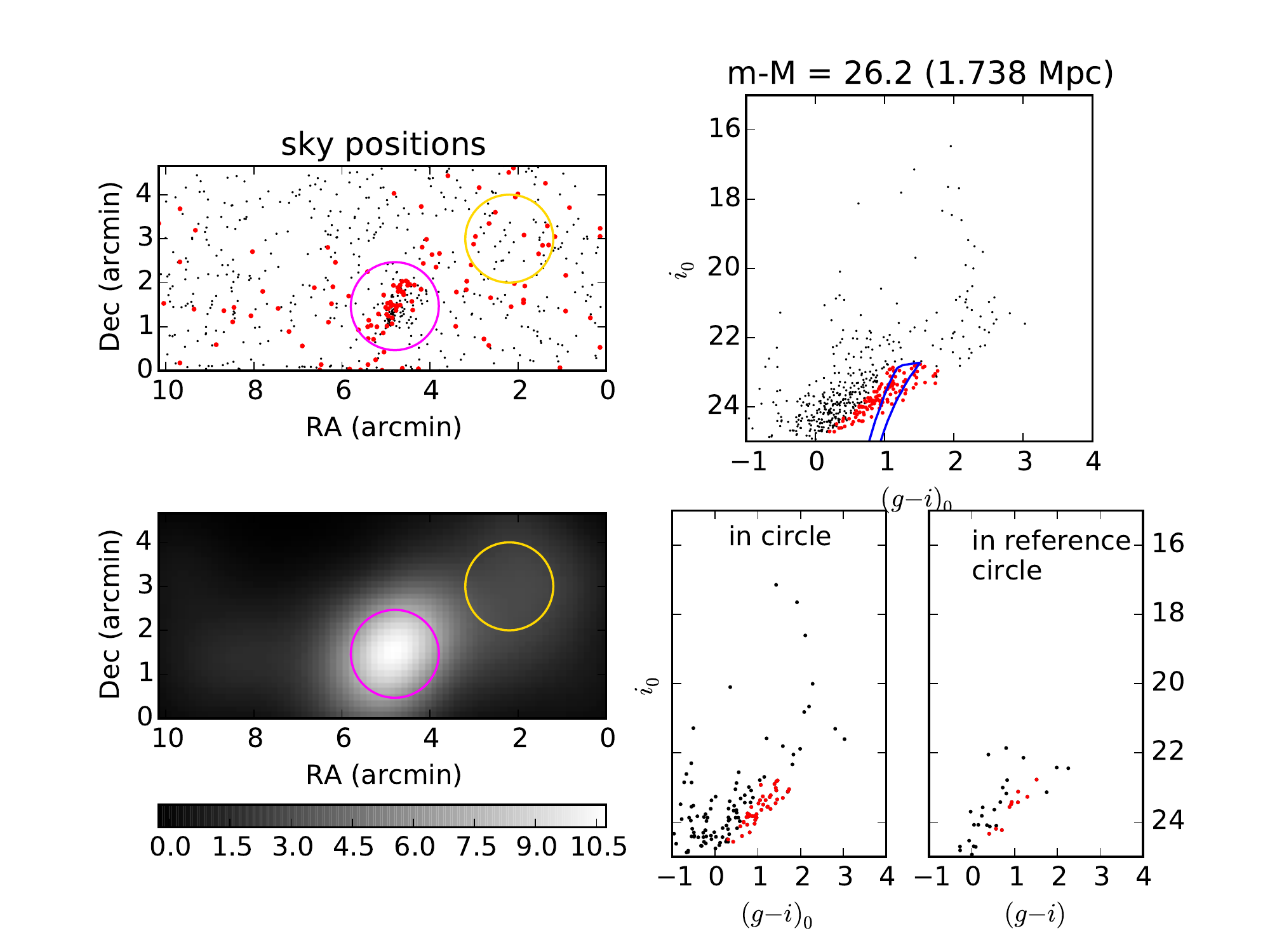}
\end{center}
\caption{Results of the filtering and smoothing process for Leo~P at a
  filter distance of 1.738 Mpc. Top left: sky positions of point
  sources relative to the corner of the field of view; small black dots are all 
  sources detected in the image, red points are
  sources in the CMD filter. A 1 arcmin
  radius circle (magenta line) is centered on the detection peak. Another 1 arcmin radius circle (yellow line) is centered at a random point to provide a reference CMD. Top
  right: color-magnitude diagram for all point sources in the field;
  the CMD filter is the solid blue line, point colors as in the top left
  panel. Bottom left: the smoothed image $S$ in units of standard deviations above or below the mean as described in Section
  \ref{section:smooth}, where whiter pixels indicate higher density; the 1
  arcmin radius circle (magenta) is centered on the highest signal
  pixel. Bottom right: color-magnitude diagrams for the stars inside
  the 1 arcmin radius magenta circle (left; colors as in other panels) and
  the stars in the yellow reference circle (right). \label{leopdetect}}
\end{figure*}

To illustrate our detection method, we apply it to images of the 
galaxy Leo~P, which was discovered by ALFALFA, originally identified as a UCHVC
based on its HI properties,
and found through optical imaging to be a dwarf galaxy in the Local
Volume \citep{leop1,rhode13,leop4}.  We analyzed the original
\citet{rhode13} images of Leo~P, obtained with WIYN and Minimosaic in
March 2012\footnote{Images of Leo P obtained with pODI in April 2013 
had poor seeing and could not be used for this analysis.}. 
Since these images were taken with broadband $BVR$
filters, the photometry measured by \citet{rhode13} was converted from
$BVR$ to $g$ and $i$ using the stellar color transformations from
\citet{sdsscolor}. After making slight adjustments to the smoothing
and significance testing methods described in Section \ref{section:filter} to
account for the difference in the field-of-view (\citet{rhode13}
analyzed a 9.6\arcmin\ by 4.8\arcmin\ portion of the Minimosaic
pointing)
we applied our filtering and smoothing technique to the photometric
measurements of Leo~P. We utilized a CMD filter that included both young and old stellar populations as well as a filter that included \textit{only} old stellar populations. Both filters produced highly significant detections; here we give the detailed results from applying the old stellar population filter for better comparison with the AGC198606 results given in Section \ref{section:198606}. We searched over the range of distance modulus given above and found an extremely strong density peak, in all cases at least 8-$\sigma$ above the mean for the field, at each distance modulus. The peak occurred at the same spatial location in the field at all distances.  In Figure \ref{leopdetect}, we show the result
of the filtering and smoothing method applied at the measured distance
of Leo~P from \citet{leop4} (distance modulus 26.20; 1.738 Mpc), which is also the most significant detection of Leo~P found with our search method.  
We begin by showing a map of detected sources in the image (the upper left panel
of Figure~\ref{leopdetect}), highlighting the sources selected by the CMD 
filtering process. We show the field CMD and filter in the upper right panel.
The source map is then binned and smoothed according to the procedure described
in Section~\ref{section:smooth}, and is shown in the bottom left panel. Here we
show a 2\arcmin diameter circle, centered on the location of the most significant 
density peak. In the bottom right panels, we show CMDs of the stars inside the circle
(putative members of Leo~P), and in a randomly placed reference circle. We see that the
CMD of stars inside the circle clearly includes a vertical feature at 
$(g-i)_0 = -1$, associated with Leo~P's blue main sequence. This is an example of the kind
of distinguishing feature we might expect to see by examining the CMD of stars near the location of a stellar overdensity, if it is associated with a dwarf galaxy with active star formation.

\begin{figure}[ht!]
\begin{center}
\includegraphics[scale=0.5]{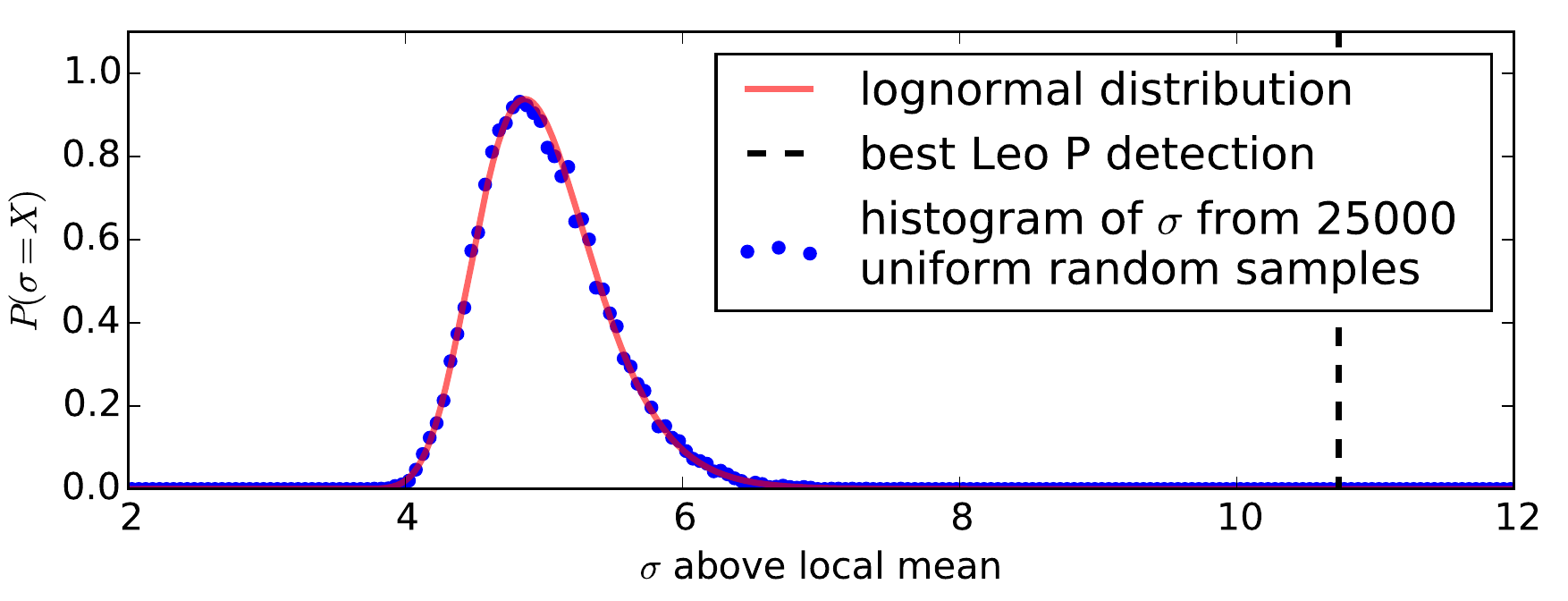}
\end{center}
\caption{Results of significance testing described in Section
  \ref{section:smooth} on the detected overdensity in the Leo~P field. Blue
  points are the histogram of density peak values for 25000 random
  distributions of 123 points. The red line is the log-normal probability
  distribution function fit to the histogram. The dashed black line is
  the location of the Leo~P detection at the \citet{leop4} distance of
  1.738 Mpc. \label{leopsig}}
\end{figure}

Leo~P is easily detected using the filtering and smoothing method. 
Figure \ref{leopsig} shows the results of the significance testing
corresponding to the peak at 1.738~Mpc and the CMD filter that includes only the old stellar population. The density peak associated
with Leo~P is an extremely strong detection at $\sim$10.6-$\sigma$, with 123 stars passing the CMD filter, causing a density peak stronger than effectively 100\% of the results from random distributions of stars
spread over a field of the same size, with the same number of stars
passing the CMD filter. These results verify that the methodology works robustly for the case of an object with a clear optical detection. We next explore how the method performs on an object without an obvious optical counterpart.

\subsection{AGC198606}
\label{section:198606}
AGC198606 is an ALFALFA-detected HI cloud with a
dynamical mass of at least $3.5 \times 10^8 d_{Mpc}$ \msun, a
rotational velocity of $\sim 14$ km s$^{-1}$, and a relatively large
size on the sky of 23\arcmin\ $\times$ 16\arcmin \citep[][]{adams15a}.  A full list of
HI-derived properties can be found in Table \ref{198606props}.  As
explained in Section~\ref{section:introduction}, this source has a radial
velocity too low to be included in the \citet{adams13} sample of UCHVCs,
but its similarity to Leo~T in terms of its HI characteristics,
location, and velocity made it an interesting target for inclusion in
our pODI imaging study.  

\begin{figure*}[ht!]
\begin{center}
\includegraphics[scale=0.8]{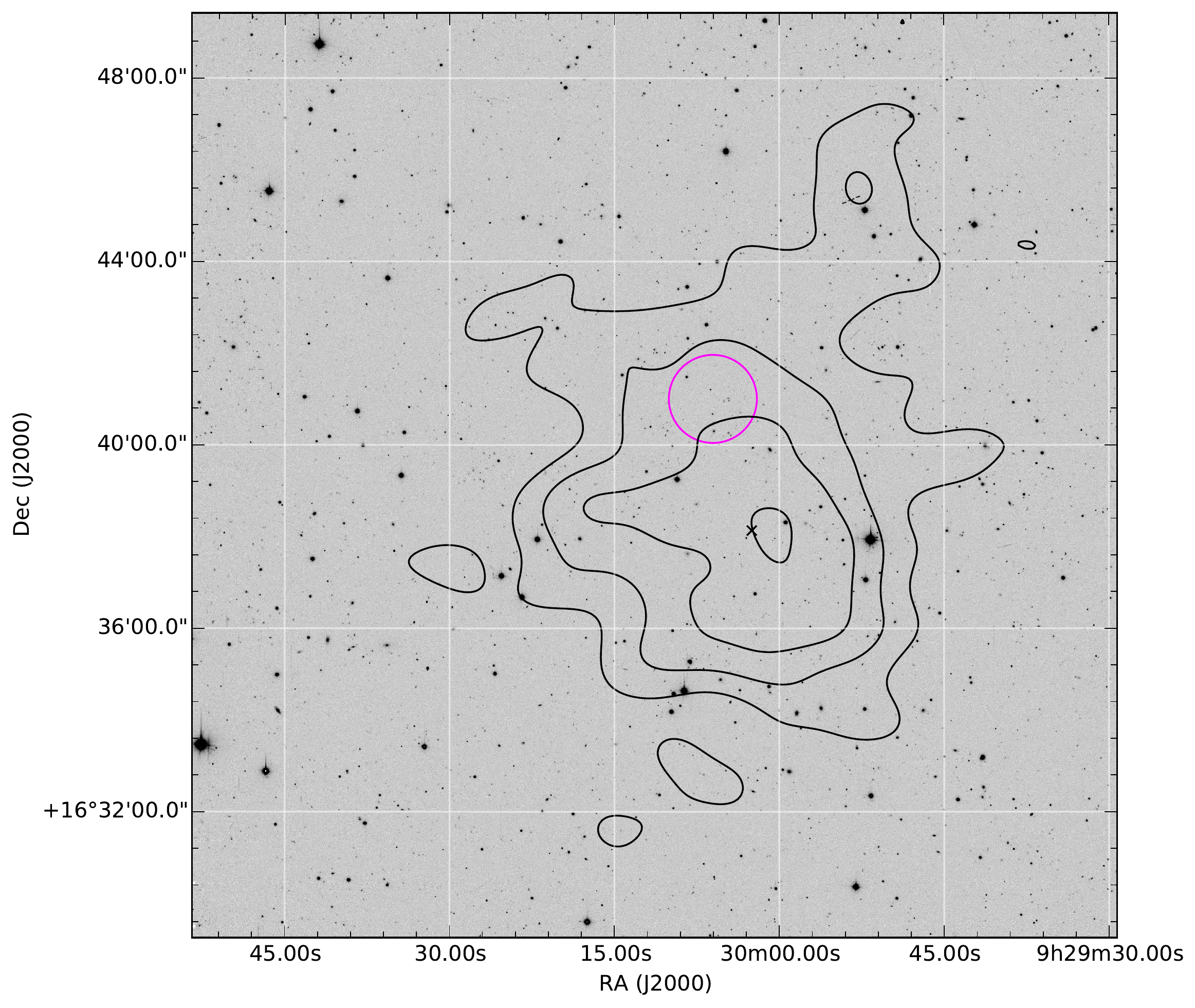}
\end{center}
\caption{WIYN pODI $i$-band image of AGC198606, overlaid with 100\arcsec\ resolution HI
  column density contours derived from WSRT observations
  \citep[see][]{adams15a}. The contour levels are [1, 2, 3, 4, 5]$\times 10^{19}$ atoms cm$^{-2}$. The location of the HI
  centroid derived in \citet{adams15a} is marked with a black cross.
  The field of view of this image is 20\arcmin\ by 20\arcmin\ and
  the orientation is N--up, E--left. The FWHMPSF of this image is approximately 0.75\arcsec. We show the location of the
  detected overdensity in the field discussed in Section \ref{section:198606}
  with a 2\arcmin diameter magenta circle.
\label{iband198606}}
\end{figure*}

Observations of AGC198606 with WIYN and pODI took place in March
2013. Integration times of 2700 seconds per filter in a nine-point
dither pattern were obtained in both the $g$ and $i$ filters. Image reduction and photometric calibration were performed in the manner described in Section \ref{section:techstuff}. Errors on the photometric zero points and color terms ranged from 0.02 to 0.05 mag. Seeing was approximately 0.75\arcsec in both filters, with a 5-$\sigma$ point source detection limit of 25.51 mag in $g$ and 24.33 mag in $i$. The final combined $i$-band image is shown in Figure \ref{iband198606} along with $100''$ resolution HI contours from \citet{adams15a}. 

\begin{figure*}[ht!]
\begin{center}
\includegraphics[scale=0.8]{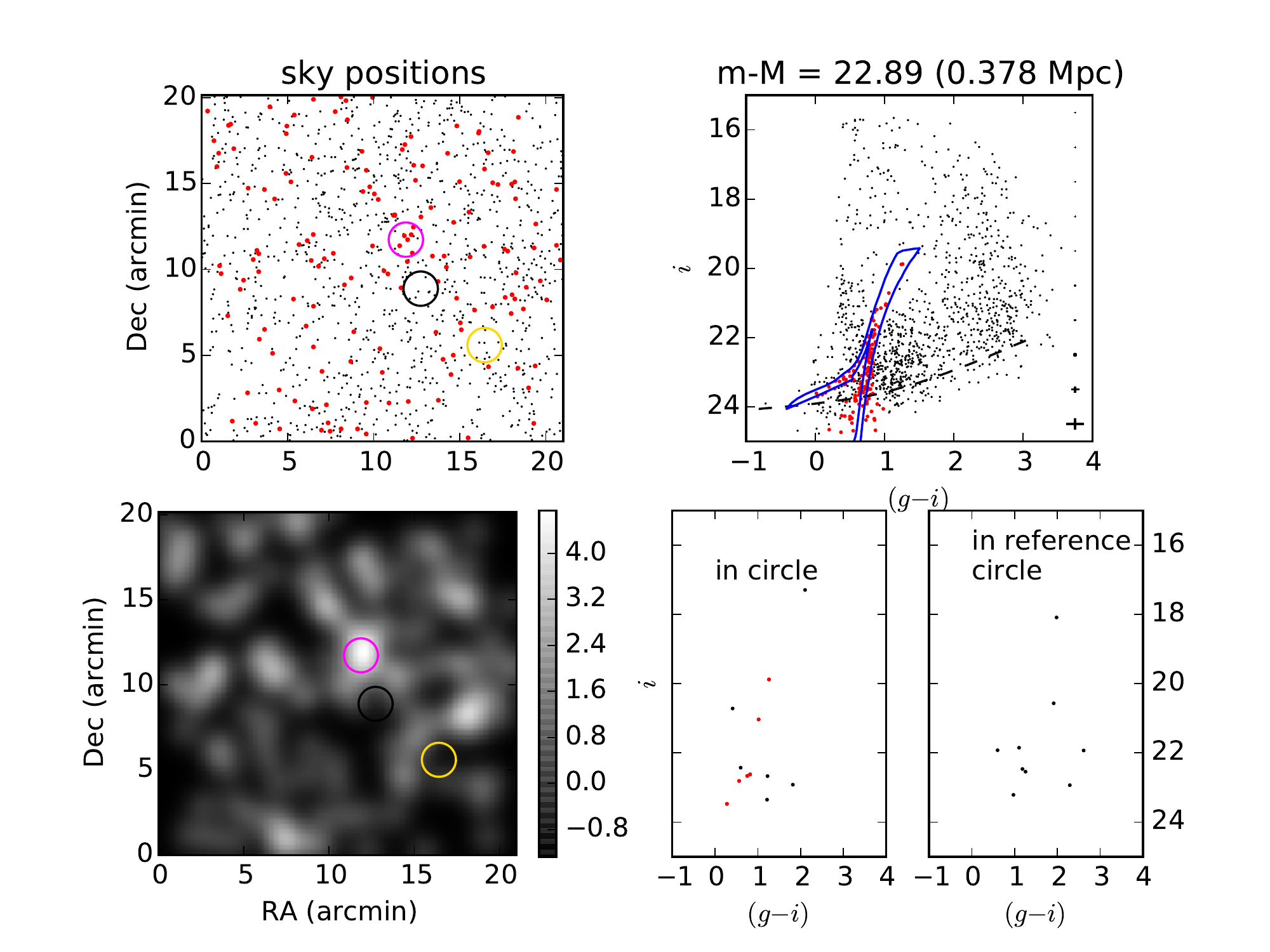}
\end{center}
\caption{Results of the filtering and smoothing process for AGC198606
  at a filter distance of 378 kpc. Panels are the same as described as
  in Figure \ref{leopdetect}. In the upper right hand panel, typical magnitude and color errors are shown along the right hand side, and the dashed black line indicates the 50\% completeness level. The 2\arcmin\ diameter magenta circle in the left hand panels corresponds to the location of the density peak discussed in the text. The 2\arcmin\ yellow circle is randomly placed and is used to select stars for a reference CMD, seen in the bottom right panel. Also plotted is a black circle
  indicating the position of the HI centroid reported in
  \citet{adams15a}.  The $\sim 4\sigma$ overdensity near the center of the
  field is a candidate for a stellar counterpart to the HI
  cloud.  \label{agc198606detect}}
\end{figure*}

We executed a series of artificial star tests to quantify the detection limits of the WIYN pODI images of the AGC198606 field.  We began by quantifying the point spread function (PSF) in the $g$ and $i$ images using a set of bright, unsaturated stars.  We added 300 artificial stars with magnitudes within 0.2 magnitude of a set value and the appropriate PSF to each of the $g$ and $i$ images.  We then performed the same set of detection and photometry steps as performed on the original images and recorded the fraction of objects that were recovered in this process.  We repeated these steps -- i.e., adding 300 artificial stars, running the detection and photometry, and recording the results -- in 0.2 magnitude intervals over a range of $\sim$5 magnitudes for each image. The  50\% completeness levels are $i = 24.1$ mag and $g = 25.1$ mag.  A dashed line in the upper right panel of Figure \ref{agc198606detect} mark the $i$ magnitude at which the convolved completeness (which takes into account the completeness level in both $g$ and $i$) is 50\% as a function of $g-i$ color.

We applied the filtering and analysis steps described in earlier
sections by shifting the CMD filter for an old stellar population over the distance modulus range described in 
Section \ref{section:filter}, from $m-M = 22.0$ mag to 27.0 mag in steps of 
0.1 mag. We found three peaks of high significance in the density of
point sources (peaks that occur less than $\sim 10$\% of the time; 
see Section~\ref{section:smooth}) that pass the CMD filter criteria 
near distance moduli of 22.4, 24.8, and 25.6, but these peaks were all 
located near the edge of the field, or outside the HI extent as shown 
in Figure \ref{iband198606}. Further inspection of these overdensities 
revealed that they were composed of faint extended sources not eliminated by 
the extended source cut and are probably background galaxy clusters. We also found an 
overdensity with $\sim 90\%$ confidence level near the center of the field at 
a distance modulus of 22.9. We more narrowly sampled the distance modulus in 
steps of 0.01 mag around 22.9, and found that the peak with 
the highest significance is associated with 
a distance modulus of 22.89 ($378$ kpc) with total of 345 objects passing the CMD filter. 
A significant peak occurred at or near the same location with a similar number of sources (344 -- 360) 
with a confidence level of 67\% or better between 
distance moduli of 22.86 (373 kpc) and 22.97 (393 kpc). We also tested the effect of the size of the smoothing scale on the detected overdensities. Larger smoothing scales do not reveal any unique strong overdensities in the image, while smaller smoothing scales typically find peak overdensities too small to be dwarf galaxies.

Figure \ref{agc198606detect} shows the results of this analysis. 
The panels in this figure are identical to those in Figure~\ref{leopdetect}, except
in the bottom left panel we indicate a 2\arcmin diameter circle centered 
on the location of the most significant density peak in the image (magenta), the location of the HI
centroid (black), and a reference circle with a randomly chosen location in 
the field (yellow). We select the stars inside the peak and reference circles and show CMDs
for each subsample of stars in the bottom right panels of the figure, allowing us to look
for evidence of CMD features present only at the location of the density peak. 
We do not use the CMD for the entire field in this case since there are a much 
larger number of stars over a much larger area than in the Leo~P field, 
and as such it would be difficult to distinguish features.
The CMD centered on the density peak in this image exhibits a largely vertical morphology, parallel
to the red giant branch from the isochrone-based filter, while the CMD from the reference circle
does not include any objects that look like red giants at any distance. The dashed black line in the upper right panel indicates the 50\% completeness level as a function of color and magnitude as described above. We also show typical instrumental magnitude and color errors as error bars along the right hand side of the upper right hand panel. Typical errors were calculated by sorting stars into ten bins based on their apparent $i$ magnitude and finding the median instrumental magnitude and color error in each bin. Values for the instrumental magnitude error are less than 0.01 for magnitudes brighter than $i=22$ and increase to $\sim 0.1$ at a magnitude of $i \simeq 24$. Color errors reach the 0.01 level at a magnitude of $i \simeq 19.5$ and the 0.1 level at $i \simeq 24$. 

\begin{figure}[ht!]
\begin{center}
\includegraphics[scale=0.5]{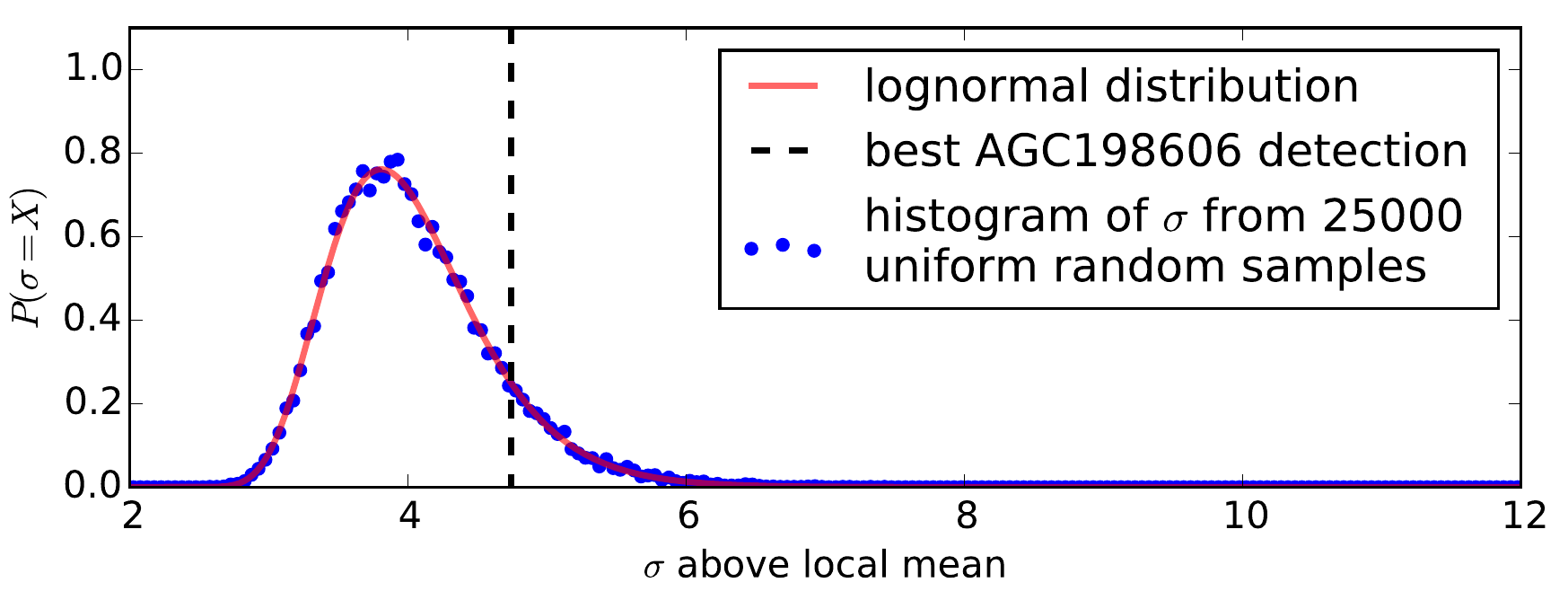}
\end{center}
\caption{Results of significance testing described in Section
  \ref{section:smooth} on the detected overdensity in the AGC198606
  field. Blue points are the histogram of density peak values for
  25000 random distributions of 345 points. The red line is the log-normal
  probability distribution function fit to the histogram. The dashed
  black line is the location of the best AGC198606 detection at $\sim
  4 \sigma$. \label{agc198606sig}}
\end{figure}

The density peak at $m-M = 22.89$ corresponding to 345 stars is $\sim 4 \sigma$ above the
local mean for this field. Figure \ref{agc198606sig} shows that the
peak is higher than 92\% of the peak values in a distribution of 25000
random samples of the same size (see Section
\ref{section:smooth}). Additionally, the peak location is only
$\sim$2\arcmin\ from the position of the HI centroid from \citet{adams15a} (indicated by the black
circle in Figure \ref{agc198606detect}) and still located within the HI disk. 

\begin{figure*}[ht!]
\begin{center}
\includegraphics[scale=0.8]{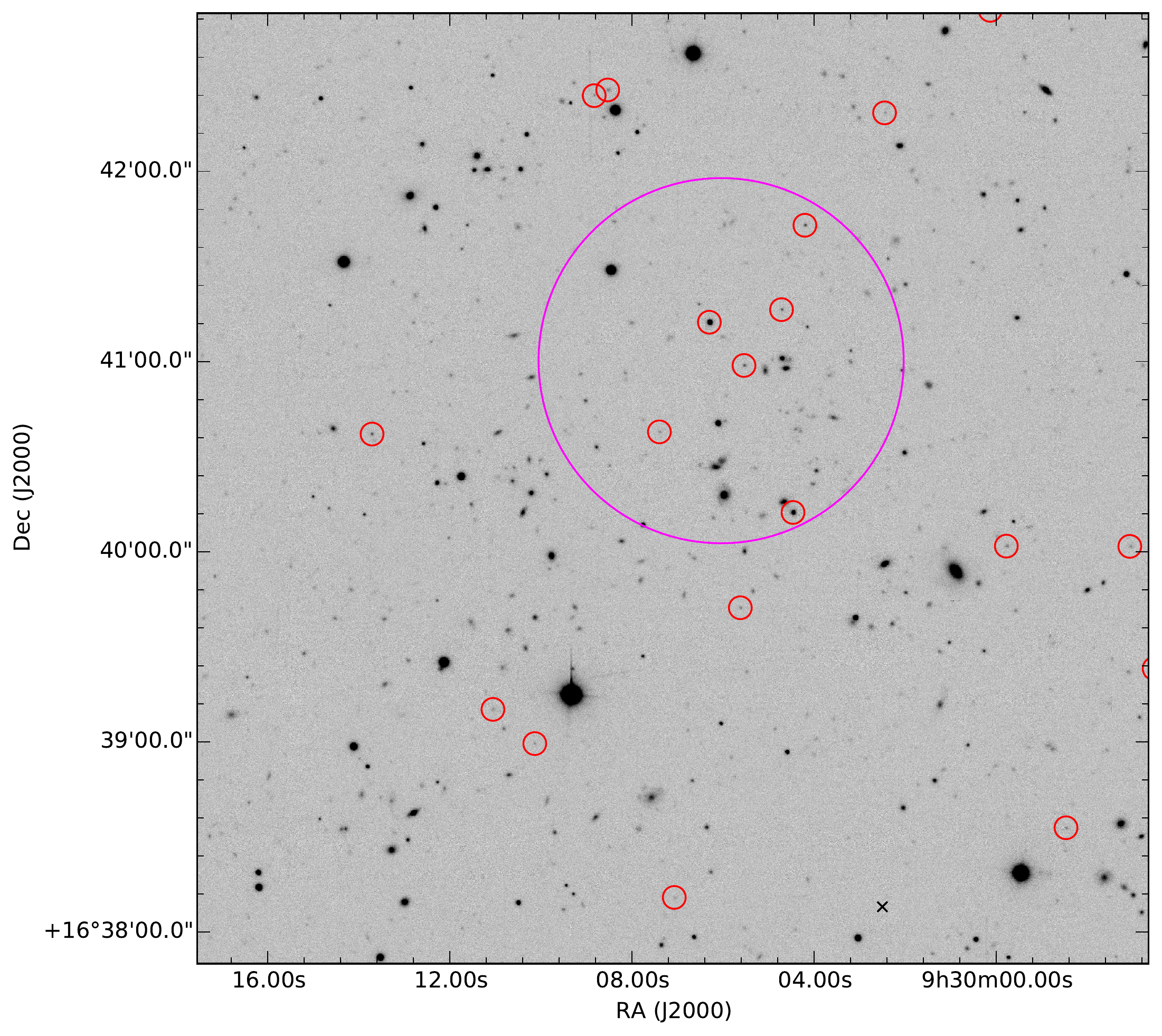}
\end{center}
\caption{WIYN pODI $i$-band image of AGC198606, shown in Figure \ref{iband198606}, cropped to a 
  field of view of 5\arcmin\ by 5\arcmin\ and centered near the density peak (marked with a 2\arcmin\  diameter magenta circle) described in the text. The orientation is N--up, E--left. The location of the HI centroid is marked with a small black cross. Stars that are included in the CMD filter are marked with red circles.
\label{5x5cutout}}
\end{figure*}

Figure \ref{5x5cutout} shows a 5\arcmin\ by 5\arcmin\ cutout of the $i$-band image centered between the density peak (outlined with a magenta circle) and the HI centroid (marked with a black cross). There is no visually apparent optical overdensity in the image,
as was noted in \citet{adams15a}. The fact that some of the detected sources in the image are very faint and could be confused with unresolved background galaxies, and the 92\% significance level of the detection 
leave open the possibility that this is not a real detection of a
stellar counterpart but is instead just the result of a chance superposition of sources with the right magnitudes and colors to pass through the CMD filter. However, the presence of some brighter stars along the red 
giant branch sequence, the location of the overdensity near the center
of the HI cloud, and the spatial and kinematic proximity of AGC198606 to Leo~T, combined with the distance at which this overdensity has been detected (378 kpc, compared to 420 kpc for Leo~T; \citealt{irwin07}), are compelling clues that this stellar overdensity may
indeed be a dwarf galaxy in the neighborhood of Leo~T. %

We can use our data to place useful constraints on the total
optical emission of the putative optical counterpart.  We performed
aperture photometry on the pODI images at the location of the detected stellar
overdensity. Stars with $i$ 
magnitudes brighter than 18 mag are brighter than the brightest star allowed in the CMD filter 
over the range of distance modulus used, and therefore are likely to be 
foreground Galactic stars and were masked in the images. We also masked red stars with $(g - i) \gapp 2.0$ 
(guided by the CMD), and obvious background galaxies in the vicinity of 
the overdensity and then measured the light within a 1.55\arcmin radius 
aperture centered on the location of the peak density. The size of
the aperture was determined by reprojecting the 1.4\arcmin half-light radius of Leo~T to 
the derived distance of the detected stellar overdensity at 378 kpc.

After correcting for Galactic extinction and doubling the measured flux to account for only measuring inside the half-light radius, we recover total magnitudes of $g=19.23 \pm 0.09$ and $i=18.22 \pm 0.09$,
with a $(g-i)$ color of 1.01. At a distance of 378 kpc as determined
above, the $i$ magnitude corresponds to an absolute magnitude $M_{i}
= -4.67$. Note that these derived quantities contain light from
unresolved sources in the AGC198606 field, 
and so the luminosity and subsequent derived values should be treated as \textit{upper limits}. Using the relationships
between stellar mass-to-light ratio and colors for \ugp passbands
presented in \citet{bell03}, we estimate the stellar $M/L$ to be 2.36,
and at a distance of 378 kpc, find an \textit{upper limit} on the stellar mass
of a potential optical counterpart to AGC198606 to be $1.2 \times
10^4$ \msun. Combined with the HI mass from \citet{adams15a}, we find a
$M_{HI}/M_{stellar} = 42.5$. 

We also derived \textit{lower limit} values for the above quantities by adding the total flux in sources that lie within the CMD filter and are located inside the 1.55\arcmin radius aperture described above. This process gives us total magnitudes
of $i=19.20 \pm 0.09$ and $g=20.18 \pm 0.09$, with a $(g-i)$ color of 0.98, and an absolute magnitude $M_{i} = -3.69$. These values give a total stellar mass of $4.6 \times10^3$ \msun\ and a $M_{HI}/M_{stellar} = 110$. The derived HI properties of AGC198606 from 
\citet{adams15a} and our derived optical parameters are listed in Table~\ref{198606props}.

\begin{deluxetable}{lrr}
\tablewidth{0pt}
\tablecaption{Properties of AGC198606 and Possible Optical Counterpart \label{198606props}}
\tablehead{
\colhead{Property}& \colhead{~} &\colhead{Value}
}
\startdata
\multicolumn{3}{c}{HI properties}\\
\tableline
R.A.  & \multicolumn{2}{r}{09:30:02.5}\\
Dec & \multicolumn{2}{r}{+16:38:08}\\
$M_{HI}$ & \multicolumn{2}{r}{{\bf $3.5\times 10^6 d^{2}_{Mpc}$ }\msun  }\\
$\theta_{HI}$ & \multicolumn{2}{r}{11\arcmin $\pm$ 1\arcmin }\\
$a\times b$ & \multicolumn{2}{r}{23\arcmin$\times$16\arcmin}\\
$r_{HI}$ & \multicolumn{2}{r}{3.3 $d_{Mpc}$ kpc}\\
$M_{dyn}$ & \multicolumn{2}{r}{$3.5 \times 10^8$ $d_{Mpc}$ \msun}\\
$N_{HI}$ & \multicolumn{2}{r}{$6\times10^{19}$ atoms cm$^{-2}$}\\
\tableline
\multicolumn{3}{c}{Estimated Optical Properties}\\
~ & bright limit & faint limit \\
\tableline
Distance & \multicolumn{2}{c}{378 kpc}\\
$g$  & $19.23 \pm 0.09$ mag & $20.18 \pm 0.10$ mag\\ %
$i$ & $18.22 \pm 0.09$ mag & $19.20 \pm 0.17$ mag\\
$M_{i}$ & $-4.67 \pm 0.09$ mag & $-3.69 \pm 0.17$ mag\\
Stellar luminosity & $5.0 \times 10^3 L_{\odot}$ & $2.0 \times 10^3 L_{\odot}$\\

Stellar mass-to-light ratio & 2.36 & 2.27\\
Stellar mass & $1.2 \times 10^4 M_{\odot}$ & $4.6 \times 10^3 M_{\odot}$\\
$M_{HI}/M_{stellar}$ & 42.5 & 110\\
\enddata
\tablecomments{HI properties reproduced from Table 1 of \citet{adams15a}. Stellar mass-to-light ratios have been estimated from relations in \citet{bell03}}
\label{table:properties}
\end{deluxetable}

\section{Discussion \label{discussion}}

The putative optical detection of AGC~198606 is clearly provocative.  For the purpose of the following discussion we are going to assume that this optical detection is real, and that the stars identified in our images do in fact represent a sparse stellar population associated with the HI gas detected as a UCHVC by ALFALFA.  We recognize that this interpretation is far from secure, but expect that future observations will be able to either confirm or refute the above hypothesis.

The detection of such a small population of stars associated with this HI cloud is consistent with models for what one might detect in a low-mass DM system that is at the boundary between systems with and without significant amounts of cold baryons.  For example, recent cosmological hydrodynamical simulations by \citet{onorbe15} of isolated dark matter halos with masses of 10$^{9.5}$ - 10$^{10}$ M$_\odot$ result in galaxies with stellar masses of 10$^4$ - 10$^6$ M$_\odot$ and large M$_{gas}$/M$_*$ ratios.  The stellar mass estimates for AGC~198606 are at the lower end of this range (10$^{3.7}$ - 10$^{4.1}$ M$_\odot$).  Simulations by \citet{br11} predict the existence of isolated ultra-faint dwarfs with similar properties.  It is possible that AGC~198606 represents exactly such a system as predicted by these simulations. 

To put the putative optical counterpart of AGC~198606 into context, we compare its properties with several other dwarf galaxies in Table \ref{DGprops}.  We include two gas-rich dwarfs (Leo T and Leo P) as well as three ultra-faint dwarfs (UFDs) that are gas-free companions to the Milky Way (Segue 1, Willman 1, and Segue 2).  The characteristics for Leo P come from \citet{rhode13} and \citet{leop4}, while those for the other galaxies are taken from \citet{mcconnachie12}.

\begin{deluxetable*}{lrrrrrr}

\tablewidth{0pt}
\tablecaption{Properties of Dwarf Galaxies \label{DGprops}}
\tablehead{
\colhead{Galaxy} & \colhead{Distance}& \colhead{$M_V$} & \colhead{$r_e$} & \colhead{$M_{HI}$} & \colhead{$M_{\star}$} &\colhead{$M_{HI}/M_{\star}$}
}
\startdata
``Friend of Leo T'' (AGC198606) & 378 kpc  & -4.7 to -3.7 & \nodata & $5.0 \times 10^5$ & $4.6 - 12.0 \times 10^3$  & 42 to 110 \\
Leo T & 409 to 420 kpc  & -8.0 & $\sim 120$ pc & $2.8 \times 10^5$ & $\sim 2 \times 10^5$ & $\sim 1.4$ \\
Leo P & 1720 kpc  & -9.4 & \nodata & $9.3 \times 10^5$ & $5.7 \times 10^5$ & $\sim 1.6$ \\
Willman 1 & 38 kpc  & -2.7 & $25$ pc & \nodata & \nodata & \nodata \\
Segue 1 & 23 kpc  & -1.5 & $29$ pc & \nodata & \nodata & \nodata \\
Segue 2 & 35 kpc  & -2.5 & $35$ pc & \nodata & \nodata & \nodata \\
\enddata
\tablecaption{Fill in appropriate citations here}
\label{table:properties}

\end{deluxetable*}

If it is indeed real, the optical counterpart of AGC~198606 moves the detection of dwarf galaxies associated with UCHVCs into a new regime.  The first discovery of an optical counterpart associated with a bona fide UCHVC was that of Leo P \citep{leop1,rhode13,leop3,leop4}, but Leo~P is both substantially further away (1.73 Mpc) and larger (M$_V$ = $-$9.4, M$_{HI}$ = 9.3 $\times$ 10$^5$ M$\odot$) than AGC~198606.   While the properties of Leo P are extreme (e.g., the lowest mass galaxy with current star formation, an abundance measurement consistent with the lowest metallicity systems known), AGC~198606 is even more extreme.  The optical characteristics of AGC~198606 are more similar to those of the UFDs discovered recently in the halo of the Milky Way \citep[and references therein]{mcconnachie12}.  While possessing a more massive stellar population than the lowest mass UFDs listed in Table \ref{DGprops}, the total stellar mass in AGC~198606 is comparable to the UFDs recently reported in the Dark Energy Survey data \citep{2015arXiv150302584T}.

When compared to its near neighbor Leo T, AGC~198606 exhibits the rather discrepant characteristics of having somewhere between $\sim$1\% and $\sim$6\% of the optical luminosity of Leo T but nearly double the HI mass.  Not surprisingly, AGC~198606 has an extreme value for the HI gas to stellar mass ratio (between 45 and 110). The HI distributions in the galaxies are rather dissimilar as well. For Leo T, the HI radius is 300 pc and the peak column density is measured to be $7 \times 10^{20}$ atoms cm$^{-2}$ \citep{rw08}, while for AGC~198606 the HI radius at the same column density level ($2 \times 10^{19}$ atoms cm$^{-2}$) is $\sim 2$ times as large (600 pc), the full HI extent is $\sim 4$ times larger ($r=1250$ pc), and the peak column density is a factor of $\sim 10$ lower ($6 \times 10^{19}$ atoms cm$^{-2}$). We note that the beam sizes used for measuring the peak column densities for the two galaxies are similar ($\sim 45^{\prime\prime}$~ for Leo T and $\sim 60^{\prime\prime}$~ for AGC~198606) but the HI extent of AGC~198606 is measured using the low resolution ($\sim 210^{\prime\prime}$) data; these mismatches should be taken into account when considering the numbers above.

Prior to the discovery of AGC~198606, Leo T appeared to be fairly isolated in space \citep{mcconnachie12}.   With a projected angular separation of 1.2$^\circ$, the \textit{minimum} physical separation of the pair is only 8.4 kpc (adopting the average distance of the two galaxies of 400 kpc).   Hence, Leo T and AGC~198606 may be separated by only a modest number of HI radii ($\sim$6 $\times$ r$_{HI}$ for AGC~198606).  More than likely their current physical separation is larger, perhaps 30-40 kpc.  However, given their similar distances and small velocity difference, it would seem plausible, even likely, that the two galaxies have had one or more close encounters in the past.  It is tempting to associate the fairly recent episode of star formation in Leo T \citep[$\sim$200 Myr ago;][]{irwin07, dejong08} with such an encounter.  The apparent lack of any obvious tidal HI bridging between the two galaxies in the ALFALFA data \citep{adams15a} would suggest that any such encounter was not a strong one. The presence of foreground Galactic emission, however, limits the constraints that can be put on any connecting emission.

It is even possible that Leo~T or its ``friend'' has a tidal origin and originated from the other.   Given the more substantial stellar population associated with Leo T, the more natural interpretation would be that such a tidal encounter with a third system might have pulled roughly two-thirds of the HI gas out of Leo T to form AGC~198606.  If so, then it would appear that any tidal model would need to be finely tuned to explain the observed system.  That is, it is hard to imagine a tidal event that could remove more than half of the gas from such a low-mass system without disrupting the system entirely.  The absence of an obvious perturber is another problem with this scenario, as is the lack of tidal HI mentioned above.  We do not favor a tidal origin for AGC~198606, although such a scenario cannot be ruled out.

As mentioned above, the optical counterpart to AGC~198606 is much more similar to the UFDs than to the more gas-rich systems like Leo T or Leo P.  It has been assumed that some UFDs represent the remnant stellar populations of heavily stripped dwarf galaxies that were accreted by the MW \citep{willman06, martin07}. The existence of the extremely sparse stellar population in AGC~198606 raises the possibility that the origin of these UFDs may have been with substantially less massive systems than has been previously supposed. That is, UFDs may have had progenitors that looked more like AGC~198606 than more massive dwarf spheroidals.  In this case, far less stripping would be required to account for the low stellar masses of some current UFDs.  While there is no reason to doubt the validity of the stripping scenario for the creation of some of the UFDs, the existence of systems like AGC~198606 provides another creation path for this class of objects.

It seems clear that additional observational work is needed before any optical detection of a stellar component to AGC~198606 can be considered secure.  The sparse nature of the putative stellar population will make a definitive detection challenging.  Two logical steps forward would be (1) deeper ground-based imaging and (2) spectroscopy of the brighter RGB stars detected in our current data.  Our current pODI images are already fairly deep and are of good image quality.  At the location of the RGB in our CMD the pODI data are 50\% complete at $i = 24.1$.  However, the location of the red clump at $i \sim 23.5$ falls where both field contamination and increasing photometric errors combine to make a definitive detection of the red clump stars uncertain.  Deeper data with reduced photometric errors would help to improve this situation.  Radial velocity measurements of the handful of possible RGB stars in AGC~198606 have the potential of confirming or rejecting their possible association with the HI gas.  Such observations were useful in the case of Leo T \citep{2007ApJ...670..313S} and are well within the reach of current ground-based spectroscopic capabilities.

Our group has obtained optical imaging data of similar depth for many additional UCHVCs selected from the \citet{adams13} catalog and from subsequent searches of the ALFALFA data by members of our team.  Initial visual searches of our current imaging data reveal no other optical counterparts as obvious as Leo P.  We are currently analyzing the data set using the automated method described here, with the expectation of either detecting additional sources similar to AGC~198606 or setting upper limits on the mass of any stellar component that might be present.

\section{Summary \label{summary}}
We have presented a processing and filtering technique designed to
detect possible optical counterparts of ultra-compact high velocity HI
clouds identified by the ALFALFA survey. 
Our filtering technique relies on constructing a region in
color-magnitude space consistent with the theoretical and
observational properties of other Local Group compact dwarf galaxies,
which we then apply to the photometry of point sources in WIYN pODI
images in order to detect overdensities in the point source maps. This
method provides an effective way to detect stellar overdensities given
appropriately deep images. Applying the technique to photometric observations of Leo~P, a
recently discovered star-forming dwarf galaxy, we recover an
extremely significant overdensity in the image, implying that less
obvious overdensities should also be detectable via our filtering and
smoothing process.

We find compelling evidence for the presence of a compact dwarf
galaxy near the location of AGC198606, a UCHVC in the vicinity of
Leo~T, the faintest dwarf galaxy in the Local
Group with recent star formation and HI gas. Our inferred properties for a potential dwarf galaxy give it a
diameter of $\sim 2$ arcmin 
at a distance of $\sim 378$ kpc, only 42 kpc less distant than
Leo~T.  An interesting candidate for follow-up study, the possible
optical counterpart to the UCHVC AGC198606 is a good
example of the capabilities of our processing and filtering methods. We are just beginning to apply the methods to the WIYN pODI imaging data we have in hand and hope to report on additional detections, or at least to set limits on the stellar populations that may be associated with the ALFALFA UCHVC sources we have targeted.

\acknowledgments
We are grateful to the staff at the WIYN Observatory, Kitt Peak, and
NOAO for their help during our various WIYN pODI observing runs. We
also thank the ODI-PPA support scientists and programmers at WIYN, IU,
and NOAO for their support and assistance with the reduction and
analysis of the WIYN pODI data. The authors acknowledge the work of the entire ALFALFA
collaboration for their efforts in support of the survey. We thank Tom Oosterloo for useful discussions, 
especially about the HI observations of this object. We would also like to thank the anonymous referee 
for reviewing the paper and providing useful comments that improved the manuscript.
W.F.J. and K.L.R. acknowledge support
from an NSF Faculty Early Career (CAREER) award (AST-0847109; PI: Rhode).
The ALFALFA team at Cornell is supported by NSF grants AST-
0607007 and AST-1107390 to R.G. and M.P.H. and by grants from the Brinson Foundation. J.M.C. is supported by NSF grant AST-1211683.

\clearpage

\end{document}